# Beyond Divergence:
# Characterizing Co-exploration Patterns in Collaborative Design Processes

Xinhui Ye , Joep Frens , and Jun Hu
Eindhoven University of Technology, Eindhoven, The Netherlands
{x.ye, j.w.frens, j.hu}@tue.nl

**Abstract**

Exploration is crucial in the design process and is known for its essential role in fostering creativity and enhancing design outcomes. Within design teams, exploration evolves into co-exploration, a collaborative and dynamic practice that this study aims to unpack. To investigate this experience, we conducted a longitudinal observational study with 61 students across 16 design teams. Over five months of weekly diary-interviews, we uncovered the intricate dynamics of co-exploration. Our main contribution is a four-dimensional framework that identifies five distinct patterns of co-exploration activities. Our findings reveal how co-exploration emerges across various activities throughout the design process, demonstrating its role in different team interactions. It fosters a sense of togetherness, keeping design teams open-minded and engaged. This engagement cultivates collective intelligence, enabling teams to actively share knowledge, build upon each other's ideas, and achieve outcomes beyond individual contributions. Our study underscores the value of co-exploration, suggesting that it reflects the trajectory of design success and warrants further research. We also provide actionable insights, equipping future practitioners with strategies to enhance co-exploration in design collaborations.

Keywords
Computer supported cooperative work
Design collaboration
Co-exploration
Design process
Diary study
Interviews
Group reflection

## Introduction

The design process is inherently iterative and exploratory, evolving through a dynamic interaction between problem and solution spaces[1], a process is often characterized as *design as exploration*[2]. Unlike the traditional *design as search*[3] process, which assumes well-defined problems and clear solutions, *design as exploration* embraces uncertainty and creativity[4]. It allows design teams to ask "what if" questions[5], explore possibilities without predetermined outcomes and continuously adapt as their understanding of both the problem and solution evolves.

Exploration is often associated with the early, divergent stages of the design process, where the aim of collaboration is to expand the pool of possibilities[6.] It is seen as a free-flowing, creative mode of working in which individuals contribute different insights, perspectives, and experiences to spark new directions and uncover opportunities[7]. When design becomes a collaborative endeavor, this exploratory nature extends beyond the individual into co-exploration. Activities such as brainstorming and ideation workshops are frequently used to facilitate this process, with the expectation that diversity of input will lead to richer possible solutions.

However, research suggests that co-exploration is not confined to the early stages of design. Wiltschnig et al. link the co-evolution of problem and solution in creative design to collaborative contexts, arguing that co-evolution is a key driver of creativity in such settings[8]. Ye et al. reported a loss of co-exploration during remote prototyping processes[9], arguing that more supporting tools should be provided for such activities in remote collaboration. While Vyas et al. observed that exploration in collaborative design occurs throughout the design process, spanning activities from brainstorming to developing interaction mechanisms to evaluating prototypes[10]. As we will show in the related work section, co-exploration appears important across multiple stages, yet it remains conceptually under-defined in both research and practice. This has prompted us to investigate it in depth. We aim to characterize co-exploration in the collaborative design process, examining its patterns, characteristics, and value. Specifically, we ask: How does "design as exploration" manifest in collaborative settings, and what mechanisms drive it? More specifically, what exactly is co-exploration? How can it be characterized within collaborative design processes? What value does it bring to the collaborative design process?

To address these questions, we conducted a five-month longitudinal observational study with 61 design students in 16 teams, combining weekly diary-interviews with a final group reflection session. Our analysis led to the development of a four-dimensional analytical framework, which we used to identify five distinct patterns of co-exploration. These findings provide a grounded understanding of co-exploration and offer practical recommendations for fostering it in collaborative design practice.

## Related work

### Collaborative design and the "co-"s terms

"Collaborative", as defined by the Oxford Dictionary, is *"involving, or done by, several people or groups of people working together"*[11]. In design, collaboration is essential because the process is complex and often requires individuals with different professional backgrounds and expertise to contribute toward a shared outcome[12]. In the literature, *collaborative design* is used in two common ways[13]: 1) as *co-design* (often used interchangeably with *participatory design*), referring to collaboration between designers and non-designers; and 2) in a narrower sense, as tightly coupled design collaboration among designers within a core design team. In a paper that examines collaborative design in depth, Kvan defines collaborative design as a close-coupled, intensive process in which participants work continuously, observing and understanding each other's moves, reasoning, and intentions[14]. In this paper, we adopt this narrower definition of *collaborative design* (type 2), focusing on the core design team that holds responsibility for the project's design decisions. This focus does not exclude co-design sessions with non-designers; rather, when such sessions meet the criteria for collaborative design, we also consider them within scope, but always framed from the perspective of the core team.

This distinction is important because it shapes how we interpret and use the growing family of "co-" terms in design research. In recent decades, terms such as co-creation[15], co-experience[16], co-making[17], and more have gained attention and undergone more investigation. However, as Mattelmäki and Visser[18] and Koskela et al.[19] have observed, these "co-" terms are widely employed across various fields without a clear consensus on their scope and definition. With that in mind, we aim to provide a clear and comprehensive explanation of our "co-" term, with a focus on understanding the value it brings to the concept of "exploration."

*(Co-)exploration in collaborative design*

Exploration is a familiar and widely used concept in design. In process models such as the *Double Diamond*, the Discover phase is explicitly framed as a period of exploring, researching, and identifying opportunities that can lead to the creation of ideas[20]. In the early *fuzzy front end*[21] of projects, design teams often diverge, intentionally expanding the problem space by fostering a creative synergy where diverse insights can collide, inspire, and build upon one another, thereby expanding the pool of possibilities[22]. Divergence is closely tied to creativity, as it is the process for generating a wide range of novel and potentially valuable ideas.

However, exploration in design is not limited to this initial divergent phase. A second, equally important use of the term refers to the iterative exploration of both problem and solution spaces throughout the design process. Fallman argues that design often involves continuously asking "what if," a mindset that pushes beyond existing paradigms and fosters innovation[23]. This perspective aligns with the concept of co-evolution[24], which highlights that design problems and solutions evolve together in a dynamic, iterative manner, requiring ongoing reflection and reframing.

When this iterative exploratory process takes place in collaborative setting, research that examines how exploration unfolds within design teams remains limited. Existing studies offer partial insights: Mattelmäki describes *co-exploring* (interchangeably with collaborative design exploration) as teams using probes approach to facilitate shared interpretations and create new understandings early in the design process[25]. Vyas et al. observe that exploration occurs throughout the collaborative design process, spanning activities from brainstorming to developing interaction mechanisms to evaluating prototypes[26]. Their study suggested that designers explore not only to address problems but also to "try out ideas, satisfy their imagination, envision and experience creativity"[27]. Ye et al. report that during the COVID-19 pandemic, teams experienced a loss of co-exploration with real prototypes and a corresponding decrease in the feeling of collaboration[28].

Taken together, these studies suggest that co-exploration encompasses a variety of designerly activities with distinct social and creative value. Yet, the literature has not fully examined in depth how such exploration functions within design teams and the specific ways it contributes to their work. This conceptual and empirical gap highlights the need for further research to define and characterize co-exploration in collaborative design processes more clearly.

# Method

## Context of the study

This study was conducted within a semester-long (20 weeks) design course at the Department of Industrial Design, xxx University (anonymized for review). The five-month study was approved by the university's Ethical Review Board (ERB) and centered on team-based design projects with open-ended briefs in two different design studios. Studio 1 focused on health and well-being in workplace environments, addressing challenges such as stress management and social interaction. Studio 2 explored connected IoT in future residences, tackling issues like balancing activities in shared spaces and visualizing energy consumption. As we previously noted, design is a co-evolving process, and students' projects follow this approach. Instead of solving predefined tasks, they worked with open-ended design briefs, first exploring the design challenge and then iteratively evolving between the problem and solution space. Throughout the course, students engaged in research, ideation, prototyping, and evaluation. Their work culminated in two group deliverables: a demonstrator presented at the end of the project and a written report detailing their design or design research process. Additionally, each student submitted an individual reflection report assessing their personal development and learning progress. During the final assessment, students were evaluated on two main criteria: the project, including the integration of relevant knowledge and the agency over the design process, and their personal development and learning progress. As these were educational design projects, students primarily collaborated within their teams without navigating complex power dynamics with external stakeholders. The students shared a design background but brought different skill sets, such as ideating, sketching, programming, and prototyping. When needed, they were encouraged to engage with external perspectives, meeting weekly with coaches and consulting potential users through field visits and interviews to refine their designs.

## Participants

We recruited participants from the abovementioned two design studios. A total of 61 participants across 16 teams took part in the study. The group included 43 second-year bachelor's students, who formed 11 teams, and 18 first-year master's students, who formed 5 teams. They entered the education program in the same year, had often taken previous courses together, and were generally familiar with one another. Each team, composed of three to four members, worked on studio-assigned tasks. The gender distribution was balanced, with 31 male and 30 female participants (self-identified). 20% of the participants were international students, with 11 out of 16 teams including international members, contributing to linguistic diversity and dynamic team compositions. Since English was the common language used for collaboration, we conducted our study in English, as all participants were able to communicate and express themselves sufficiently. We used a meeting room next to their open working space, making it convenient for participants to bring relevant artifacts for demonstration during interviews. Note that while all students completed the same deliverables, assessment expectations differed between bachelor's and master's students to reflect their respective learning outcomes.

### *Mixed methods in this research*

We used a mixed-methods approach, combining qualitative and quantitative methods to understand co-exploration over a five-month longitudinal study. Figure 1 provides an overview of our two-stage study (left: study sequence; right: analysis sequence).

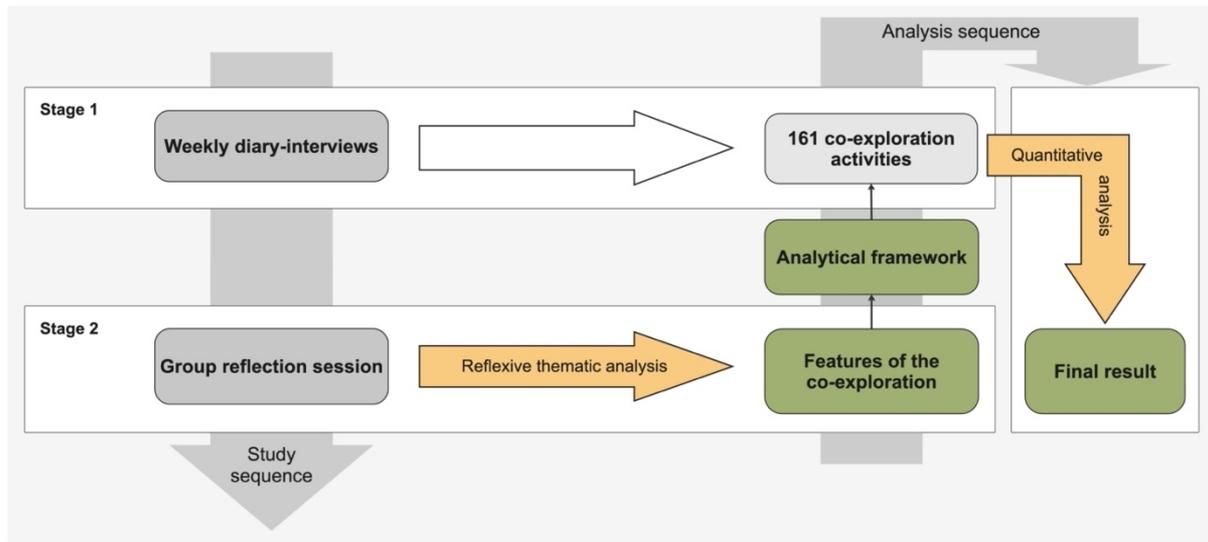

Figure 1. Overview of the study procedure. Left: a two-stage study including weekly diary-interviews and a group reflection session. Right: reversed order of data analysis compared to the sequence of data collection

Stage 1 involved weekly diary-interviews. Each week, participants met with the PI (first author) to collaboratively document both individual and team design activities from the past week on a shared Miro board. These diaries included photos, videos, links, transcriptions, and field notes. Following this documentation of diaries, the first author conducted semi-structured interviews, prompting participants to walk through their entries and elaborate on the week's activities.

Stage 2 was a group reflection session at the end of the design project. Here, participants, as experts of their own experiences[29], revisited their co-exploration activities, clustered them, explained their reasoning, and reflected on what co-exploration meant to them.

On the analysis side (right of Figure 1), the process followed a reversed order from data collection. While the diary-interviews captured rich, ongoing details, they lacked the comprehensive perspective of hindsight that is needed to contextualize actions fully. In contrast, the group reflection session provided a holistic view, enabling participants to articulate their understanding of co-exploration. We therefore began analysis with the group reflection data, using reflexive thematic analysis[30] to identify shared *features* of co-exploration. These features informed the development of an analytical framework, which we then applied to the diary-interview data to code 161 co-exploration activities. Using this coded dataset, we employed parallel coordinate plots[31] to analyze and visualize activities into five distinct patterns. Finally, after the project concluded, we collected participants' final assessment results and combined them with our observational assessments. We conducted descriptive and correlational analyses to explore potential relationships between the type, frequency, and variety of co-exploration patterns, the design stage (timeline), and the design outcomes.

In the following sections, we detail the study procedures and analysis methods.

## Data collection approach

### Stage 0: Kick-off meeting

We conducted a kick-off meeting to introduce the study, schedule recurring time slots for weekly interviews, and obtain informed consent. This session clarified study procedures, assured participants that the PI had no role in their assessments and emphasized their right to withdraw at any time. Participants were given full autonomy in choosing how they collaborated - whether face-to-face, hybrid, or fully online. However, the PI expressed a preference for conducting the weekly interviews face-to-face when possible. The PI then guided participants through the diary template they would use. While reviewing the template, participants encountered questions on co-exploration, prompting a brief explanation. The PI explained that a decline in co-exploration has been observed in remote collaboration, emphasizing the need to investigate how design researchers can better support this process[32]. Given the absence of a working definition, the study invited participants to document and interpret co-exploration as they experienced it. The meeting concluded with an open discussion, addressing any remaining participant questions.

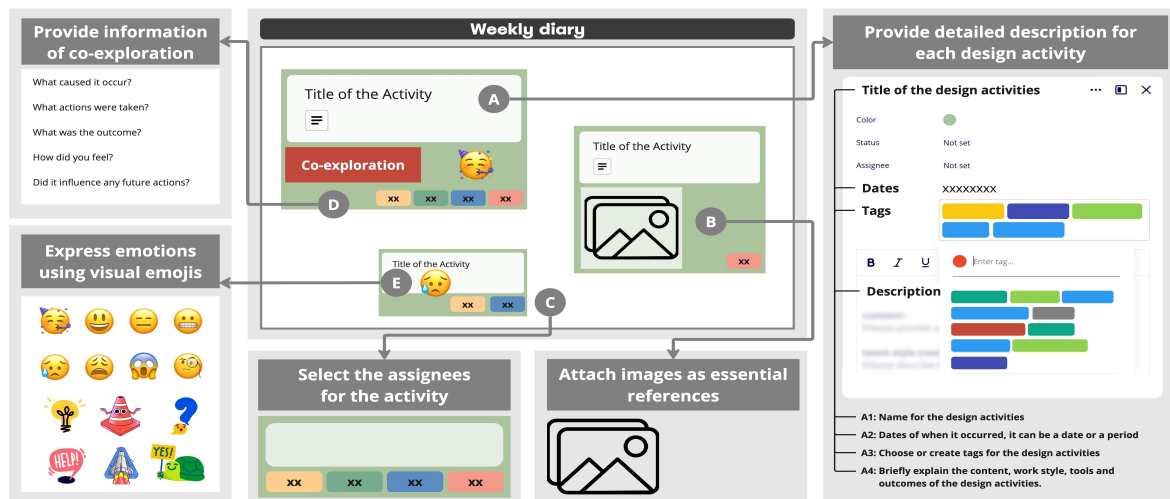

Figure 2. A schema of the weekly diary entry. Participants began with (A) providing a detailed description for each design activity, then (B) they were encouraged to attach images as references for later discussions. After (C) selecting assignees for each activity, participants (D) identified the activities as co-exploration and explained their reasoning. Finally, they (E) expressed their emotions using emojis for insights beyond text descriptions

### Stage 1: Weekly diary-interviews

In Stage 1, each design team participated in a weekly diary-interview session. This involved two linked activities:

---

1. Completing a shared group diary to document both group and individual design activities from the past week.
2. A follow-up group interview in which the diary entries were elaborated and discussed.

Diary entries[33] are a valuable method for observing participants in design research. Advances in digital tools have made image and audio diaries more accessible, and the diary-interview method offers deeper insights into participants' evolving experiences over time[34]. For our study, we used Miro for weekly diary entries because: (1) it was a familiar platform for students from the COVID-19 period, allowing multiple users to fill in the diary simultaneously, and (2) it enabled students to digitize their design materials throughout the entire process, benefiting their final reports, which were a required deliverable for their assessments. The diaries were completed by the students themselves, either before the interview or during the first part of the session. While they filled out the diary, the PI remained silent, allowing them to discuss and document activities collaboratively.

Once the diary was complete, the PI conducted a semi-structured group interview, prompting participants to walk through their entries, elaborate on details, and reflect on the week's activities. During the interviews, the PI listened, occasionally asking questions to ensure that the complete weekly design process diary was discussed. Turn-taking was carefully monitored to ensure every team member had the opportunity to speak and contribute their perspectives. To capture additional details, field notes were taken immediately after each interview, documenting communication styles, connections to prior entries, team dynamics, and other observations. The interview was audio-recorded (with consent) and supplemented by field notes capturing team dynamics, communication styles, and links to prior entries. We used Microsoft Teams for the interviews because: (1) it provided live transcription, ensuring an accurate record of discussions for later analysis, and (2) it allowed remote participation for those unable to attend the weekly diary-interview sessions in person, ensuring full data collection.

As shown in Figure 2, participants followed a structured weekly protocol on the Miro board. Participants began by initiating a new diary board, which contained templates for documenting activities. They were instructed to provide detailed information for each design activity, including the title (A1), date (A2), tags related to different actions (A3), and a brief description (A4). They were encouraged to attach images as references during explanations and to serve as prompts in later interviews (B). Assignees for each activity were selected by participants, with name labels easily removable if certain individuals did not participate (C). Next, participants were asked to share which design activities were perceived as co-exploration and their selection reasons (D). Finally, participants expressed their emotions about activities using visual emojis, providing insights beyond textual descriptions (E). Note that the diary board was continuous, allowing participants to revisit and edit previous entries and link related activities. This structure, as seen in Figure 3, enabled a clear view of the design process, with connections indicated between activities, such as linking individual sketches to group discussions.

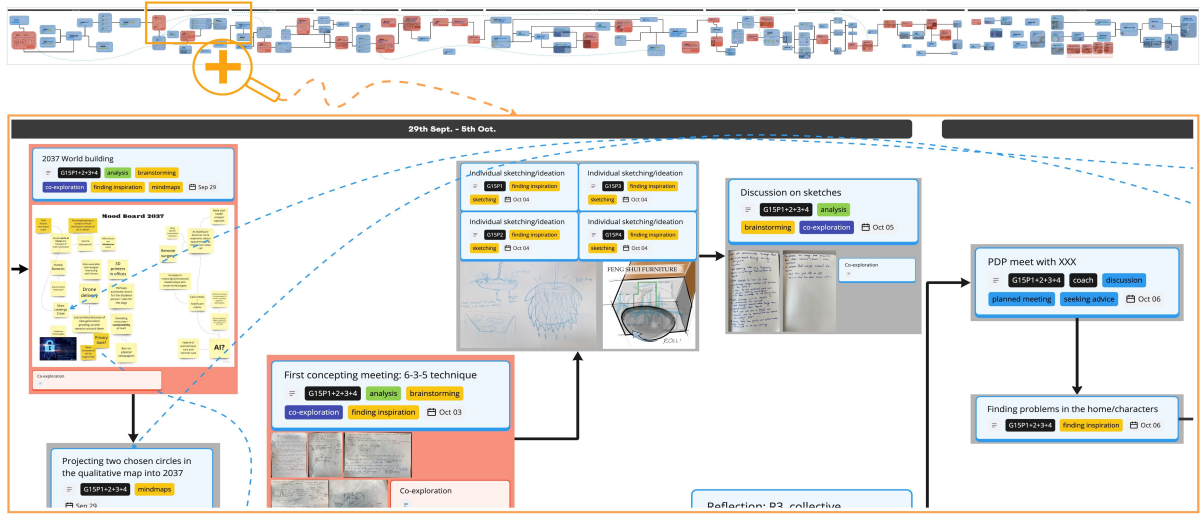

Figure 3. Overview of team's design process and a close-up of their diary documentation: world-building for a design challenge (leftmost section), brainstorming with the 6-3-5 technique, assigning sketching tasks (middle section), and group meetings (rightmost section).

### Stage 2: Final group reflection session

At the end of the design project, a final group reflection session was conducted, using a card-based tool. Such tools have been widely used in design activities to facilitate reflection, ideation, and inspiration[35]. For instance, *Method Cards* help designers empathize with users in design projects[36], the *Intent Toolkit* facilitates collaborative ideation during brainstorming[37], PLEX Cards used for inspiration when designing for playfulness[38]. In educational settings, *Progress Cards* have supported design students in reflecting on their processes[39]. Inspired by these tools, we adapted this approach for the final group reflection session. We transformed digital descriptions of co-exploration activities from the interviews into physical cards, as depicted in Figure 4. Each card featured digital content from the weekly diaries on the front, serving as a tangible representation of their design activities. While the back side featured key points and quotes to prompt participants' recall of details. The group reflection session began with a sensitizing exercise, where participants clustered the cards by similarity, reading, organizing, and interpreting their co-exploration activities retrospectively. This hands-on interaction refreshed their memories and deepened their understanding of their co-exploration activities, setting a solid foundation for addressing the interview questions about the definition of co-exploration:

Q1: How do you define co-exploration in a design team?

Q2: What characterizes an activity as co-explorative?

Q3: What does co-exploration mean to you?

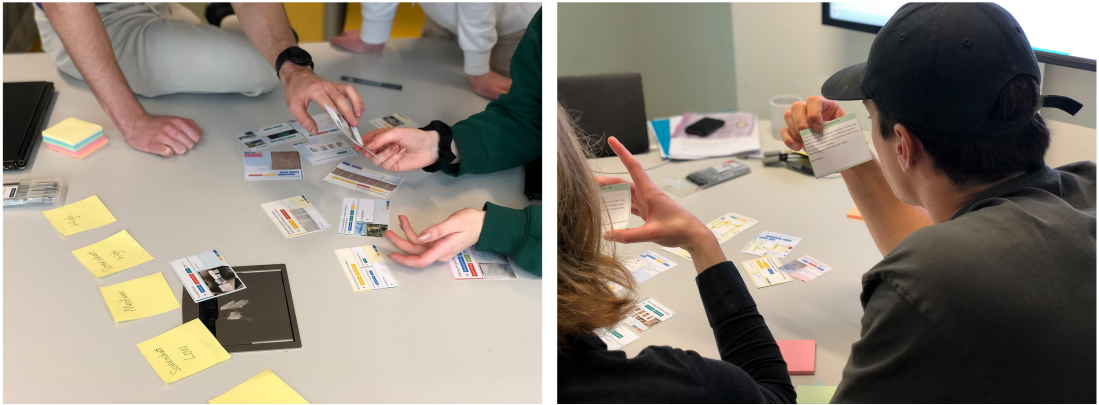

Figure 4. Participants engage in evaluating their co-exploration activities. Left: Participants engaging in evaluating their co-exploration activities. Right: Participants read the back side to recall the details of their co-exploration activities.

## Data analysis approach

As introduced in Section 3.3, our analysis followed a reversed order compared to data collection. We started with the final group reflection session data to establish a broad understanding of co-exploration, which set the foundation for the development of our analytical framework. This framework was then applied to analyze specific activities from the diary-interviews, allowing us to map and categorize co-exploration patterns. Below, we detail our analysis approach.

### *Analysis of group reflection session*

We transferred responses about co-exploration into MAXQDA 2022 and involved a second coder from our department for qualitative analysis. Following the reflexive thematic analysis approach outlined by Braun and Clarke[40], our analysis proceeded as follows: We (1) familiarized ourselves with the data by reading it multiple times, and to ensure a shared understanding, we provided the second coder, who was not involved in the interviews, with all the diaries. (2) We then generated initial codes and (3) subsequently organized them into themes. (4) We collaboratively reviewed and discussed our respective codebooks, with the goal of identifying shared themes. (5) We reached an agreement on the themes and went through the data one more time. (6) As we prepared this paper, the themes were further refined, resulting in the following five features of co-exploration.

The findings from the group reflection session are presented here immediately after the analysis methods because they form the foundation for the subsequent analysis: Although these findings are part of the results, they were crucial for developing the analytical framework. We believe that presenting them here ensures a logical flow, providing coherence between the stages and clarifying how the insights from the group reflection informed the diary analysis.

---

[40] Braun, Virginia, and Victoria Clarke. "Using thematic analysis in psychology." *Qualitative research in psychology* 3, no. 2 (2006), pp. 77-101.

Being "together" is key

The term "together" was mentioned 76 times and emerged as essential in co-exploration. It comes as no surprise that togetherness holds a prominent position when defining concepts prefixed with "co-" in the field of collaborative design. We examined participants' descriptions to understand what togetherness means to them and how it is expressed and interpreted in co-exploration activities.

*Together, in what sense?* "Together" can refer to physical co-presence or a sense of togetherness, even when individuals are not physically together. T4P4 (Team 4, Participant 4) and T5P4 highlighted the importance of physical proximity, such as "sitting together" or being present in person, emphasizing the need for co-presence. However, some participants, such as T3P2, expressed that "online/offline does not matter," pointing to a sense of togetherness that transcends physical location.

*Together, with whom?* Participants from teams T1, T7, T9, T11, and T13 discussed interactions within their groups, underscoring the collaborative nature of their design work. T1P4 expanded this to include "people outside of the project," suggesting that co-exploration could extend beyond the design team to external collaborators.

*Together, doing what?* Participants described various collaborative activities. "Discussing" was the most frequently mentioned, cited 37 times by 14 groups. They also referenced activities such as "brainstorming," "ideating," and "sketching together." From these descriptions, it appears that co-exploration occurs across a wide range of design activities. However, we have yet to investigate what transforms a set of activities into a co-exploration experience.

The synergy of different insights

Forty-five participants stressed the value of synergizing diverse inputs, such as different ideas, knowledge, and perspectives. T12P2 highlighted the need for everyone to contribute, while T5P3 noted that co-exploration could start with individual findings. Additionally, 6 participants (T3P4, T4P4, T6P1, T8P2, T14P1, T14P2) mentioned drawing on past experiences and leveraging each other's skills to enrich the co-exploration process.

Being open-minded and active

Sixteen participants from seven teams highlighted the importance of building on each other's ideas, using descriptions like "bouncing off each other's ideas" (T15P2) and engaging in a "ping-pong manner" (T2P4). T4P1 described co-exploration as an "active push." To facilitate this dynamic exchange, T7P2 noted that group members should freely express themselves, while T15P1 stressed the value of everyone being heard and actively listened to. T15P3 emphasized the need for equal participation, with all members having "an equal seat in a meeting." T13P4 demonstrated constructive critique by suggesting "nice idea, and we can..." instead of the limiting term "but."

Different types of communication

We identified four types of co-exploration: 1) Sharing and exchanging, mentioned by 10 participants across 8 teams, focused on gaining "new perspectives" (T8P3&P4) and learn "new knowledge" (T10P4, T14P1). 2) Improving, highlighted by 15 participants from 11 teams, involved "deepening" (T10P1, T16P2) and "broadening" (T5P1, T9P1) concepts. 3) Combining, noted by 7 participants from 5 teams, included activities like "making decisions" and "propelling the design process" (T2P2, T7P1, T16P3). 4) Open-ended communication, acknowledged by 5 participants, emphasized that co-exploration "does not always need to directly result in conclusions" (T4P4, T8P2), instead fostering shared understanding. This sentiment was expressed as "my brain is someone else's brain too" (T15P2), and described as "combining beliefs" (T3P3).

Social and functional values of co-exploration

Participants highlighted both social and functional values of co-exploration. T16P4, T3P1, and T11P4 underscored its importance for team building, saying, "it's important to me [...] it helps groups become more of a team" (T11P4). Trust was also enhanced, as noted by T6P3 and T16P4. Functionally, T8P4 described the process as yielding outcomes that are "more than the sum of individual capabilities," a sentiment echoed by eight participants in eight teams. Furthermore, "high efficiency" (T1P3, T2P2) and "better solutions" (T1P4) were also cited as critical benefits of co-exploration.

### *Analysis of weekly diary-interviews*

Framework development

To identify patterns in co-exploration across different design topics and team dynamics, we need to use the Parallel Coordinate Plot (PCP) method [41], which facilitates the exploration and visualization of multidimensional relationships. PCP has been widely used in design research to uncover patterns and insights, such as identifying design patterns for ambient information system[42], analyzing different card-based design tools[43], and mapping designs to frameworks to identify common usage patterns[44].

For PCP to be effective, the data must be described using well-defined dimensions with clearly specified values. The features of co-exploration identified in the group reflection sessions were valuable for recognizing when co-exploration occurred, but they were not immediately usable for PCP analysis. They were too broad and, crucially, they lacked clearly defined, countable values that would allow activities to be categorized and compared systematically. At the same time, these features already captured important aspects of co-exploration as experienced by participants, and thus provided a meaningful foundation. Therefore, rather than discarding them, we chose to refine the features, translating them into dimensions with discrete values that could be consistently coded in the diary–interview data. We refined the features using two criteria:

1) Categorization potential: Could the feature be expressed as a dimension with clearly defined values that distinguish between different co-exploration activities?
2) Data connection: Could the dimension and its values be directly observed and coded in the weekly diary-interview data?

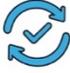

Figure 5. Four dimensions for characterizing co-exploration activities.

The following describes how each feature was refined into a dimension:

1. The feature Being "together" is key described both physical co-presence and a sense of togetherness in online collaboration. While this was valuable for recognizing co-exploration, it was too general to categorize activities in a way that would reveal patterns. Because co-presence can also be experienced digitally, we emphasize *physical* co-presence when referring to situations where team members were in the same location. We therefore reframed this feature as the People Distribution dimension, specifying values that captured the different configurations observed in the diary-interview data: physically co-present, hybrid (hand over), hybrid (some joined online), and completely online.

*2.* The synergy of different insights captured the value of combining diverse ideas, knowledge, and perspectives, but did not specify how these insights were generated. To make it more descriptive and directly observable, we refined it into the Diversity of Insights dimension, with three values: *Individual preparation* (insights were prepared in advance), *Generated by group-based design techniques* (insights were created collaboratively through methods like brainstorming, role-playing, or mind-mapping), *Knowledgeable with no preparation* (participants relied on each other's existing knowledge and skills without prior preparation).

3. The feature Being open-minded and active, while important as an attitude for co-exploration, could not be expressed as a concrete, observable dimension with distinct values. It described a quality of interaction rather than a characteristic of the activity, and therefore did not meet our criteria for inclusion in the framework.

4. The feature *Different types of communication* was already closer to a categorization, but the participant language (e.g., sharing and exchanging, improving, combining, open-ended communication) needed to be aligned with terminology more commonly used in design research. We therefore refined it into the Communication Types dimension, with four values: *Converging: combining* (integrating insights into unified concepts), *All-sides refining* (collective refinement without merging), *One-side refining* (enhancing one participant's idea) and *Diverging: generating more* (open-ended brainstorming to generate diverse ideas). This rephrasing preserved the meaning of the original categories while making them more precise and analytically consistent.

5. Finally, during the process of linking features to the diary-interview data, we identified an important factor not captured in the original features: the way information was distributed prior to an activity often shaped how the co-exploration unfolded. This became the Information Distribution dimension, with three values: sync (participants had fully shared information before starting), context aware (participants had partial awareness of others' activities), and async (no information shared beforehand).

The resulting analytical framework (Figure 5) contains four dimensions: Information Distribution, Diversity of Insights, Communication Types and People Distribution. These dimensions and their values provided the structured coding scheme needed for PCP analysis, enabling us to analyze co-exploration activities across different teams' collaborative design processes, and to identify recurring patterns in how co-exploration occurred in practice.

Applying the analytical framework to the data

To illustrate how the analytical framework was applied using the PCP approach, the left side of Figure 6 shows the categorization of one example activity: First Concept Meeting: 6-3-5 Technique (previously introduced in the close-up diary in Figure 3). In this brainstorming session, designers first sketched ideas individually without prior sharing, which we coded as *Async*. The use of the 6-3-5 technique, a structured *group-based design technique*, meant that ideas were generated collaboratively, leading to the *Diverging: generating more*. Since all participants joined the brainstorming in real person, the final dimension was marked as *physically co-present*. As shown in Figure 6, mapping these four dimensions onto the framework provides a structured visualization of this activity and demonstrates how it fits within the broader analytical framework.

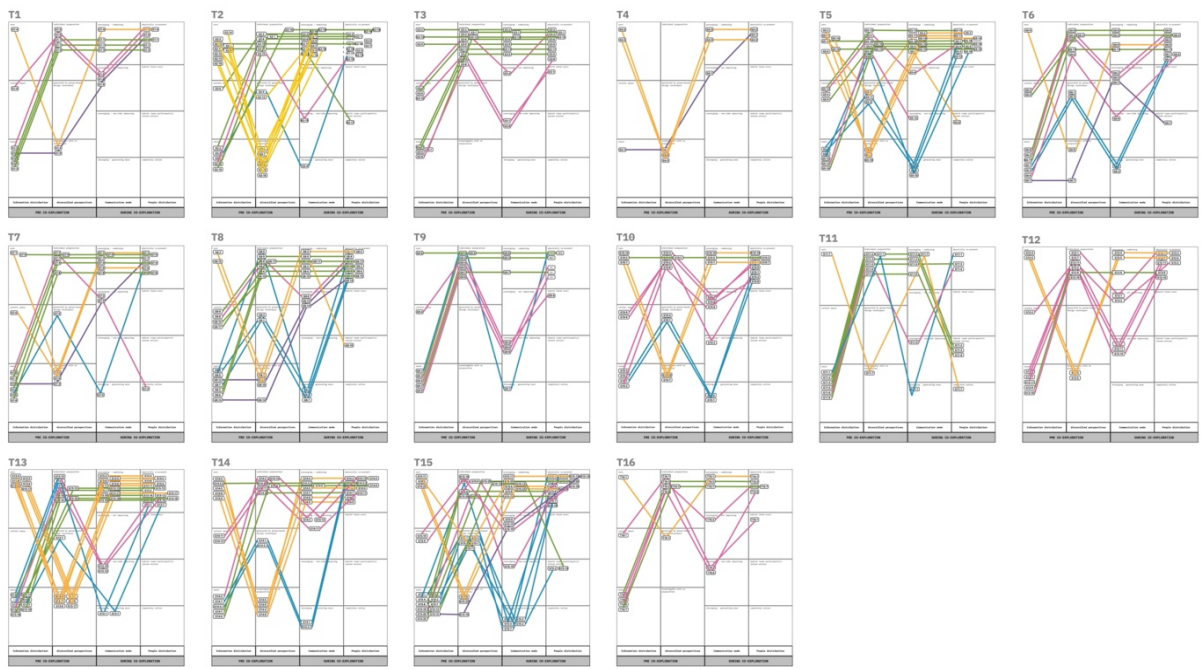

| Sync | Individual preparation | Converging combining | Physically co-present |
| | | | **T15-3** |
| | | All-sides refining | Hybrid (hand over) |
| Context aware | Group-based design techniques | | |
| | **T15-3** | One-side refining | Hybrid (some joined online) |
| Async | Knowledgeable with no preparation | | |
| **T15-3** | | Diverging generating more | Completely online |
| | | **T15-3** | |
| Information Distribution | Diversity of Insights | Communication Types | People Distribution |
| **PRE CO-EXPLORATION** | | **DURING CO-EXPLORATION** | |

Figure 6. Using "First Concept Meeting: 6-3-5 Technique" - the third recorded co-exploration activity from Team 15 - as an example, the PCP approach analyzes co-exploration activities.

Using the same process, we analyzed all 161 co-exploration activities, coding each one according to the four framework dimensions and their values. Figure 7 presents an overview of 16 analytical frameworks, one for each design team, where each line represents a unique experience of co-exploration activity. Different colors distinguish patterns of co-exploration, as further elaborated in the results section.

Figure 7. An overview of 161 co-exploration activities from 16 teams in the analytical framework. Each line represents a unique co-exploration activity. Different colors distinguish groups of co-exploration activities with similar patterns, as further elaborated in the result sections.

## Results

The study yielded a rich dataset from the weekly diary-interviews, including 1562 documented activities and over 78 hours of audio material (details in the Appendix), with each session averaging about 29 minutes. Among all the design activities, 220 activities across 16 teams were initially tagged as co-exploration during the diary-interviews, with participants explaining their reasons for the categorization. Due to the varying richness of documentation, our analysis focused on 161 well documented co-exploration activities. Figure 7 presents these 161 activities mapped across their dimensions and values, revealing five distinct patterns (over all groups). To comprehensively illustrate these patterns, we used heatmaps, with cell colors indicating the frequency of cases associated with each value.

In the following sections, we first present these five patterns of co-exploration. Each pattern is illustrated using typical episodes from the data and is defined by a summary of its shared characteristics. After this qualitative presentation, we report the results of our quantitative analysis, which examined differences across teams regarding the frequency and diversity of their co-exploration activities.

### *Co-exploration when merging different insights*

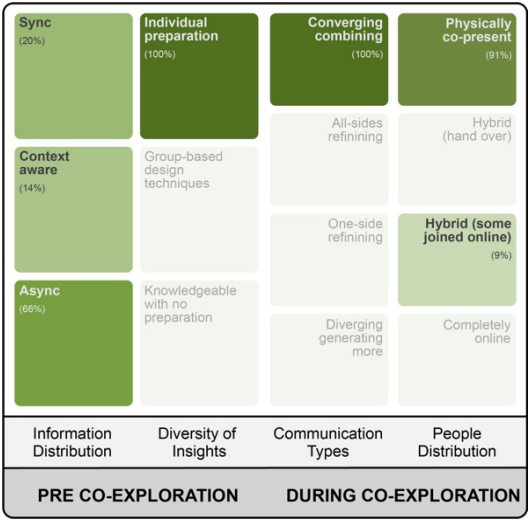

Figure 8. Co-exploration when merging different insights

Episode 1: *"Group meeting: Merge!!"*
During a prototyping day, Team 8 generated multiple lo-fi prototypes, each reflecting different ideas. Due to time constraints, they reconvened to further discuss their prototypes two days later. As described by T8P1, they wrote down "merge!!" on the whiteboard, signifying the pivotal moment when they decided to consolidate their ideas into a single concept. T8P2 explained, *"At the start, everything was still vague. We all had ideas [...] but we knew that we need to have one, just one."*

This pattern unfolds throughout the entire design process and is underscored by two key characteristics. First, individual preparation was impactful whether ideas were shared beforehand (20%), contextually aware (14%), or entirely independent (66%). The second characteristic is "converging- combining." Methods like voting, ranking, and detailed discussions enabled teams to merge insights effectively. Regarding people distribution, 51 co-exploration activities occurred when all team members were physically co-present, with only five cases involving partial online participation. Participants reflected on the impact of this pattern during group reflection sessions, frequently describing it as having a "high impact" and driving "breakthroughs in the design process," reinforcing its perceived "high importance" in shaping their collaborative outcomes.

### *Co-exploration when reflecting on existing materials*

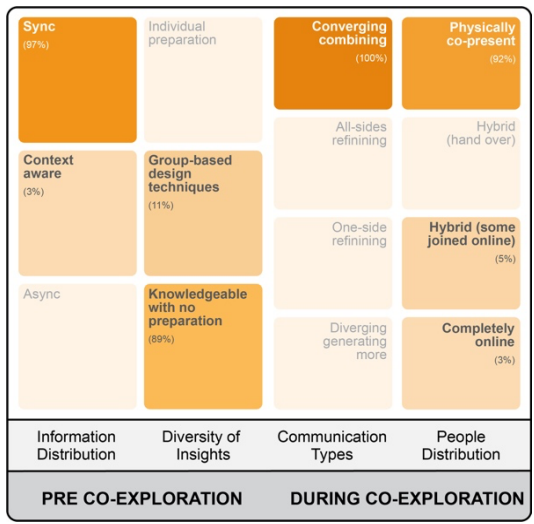

Figure 9. Co-exploration when reflecting on existing materials

Episode 2: *"Discussing concerns and thinking of ways to solve it"*

After the mid-term demo day, Team 7 faced difficulties refining their concept while other teams had already started prototyping. They presented their individual preparations and attempted to merge them into one concept. However, things remained apart. P3 mentioned, *"We made some decisions that I'm not sure about. I slowly felt distance from the project. I felt lost."* During their meeting, team members expressed concerns, and P1 later noted, *"We had a better overview of the framework of our project, and we were all on the same line again now."*

This pattern, observed in 38 activities, focused on aligning team understanding. It relied on shared or contextually aware information and often resulted in teams "getting on the same page," either by making concrete decisions or achieving a shared clarity of objectives. To navigate this ambiguity, discussions often involved visual aids like mind maps, and occasionally, group-based design techniques like role-playing and co-sketching. A key insight from cases like Episode 2 is that confronting ambiguity could evoke frustration, with tension, and even tears observed during some session. These emotional moments were not necessarily signs of team dysfunction, but rather part of the difficult work of negotiating meaning when clarity is lost. Ultimately, these challenging co-exploratory sessions led to breakthroughs, creating the shared clarity necessary to realign perspectives and advance the project collaboratively.

### Co-exploration when refining initial concepts

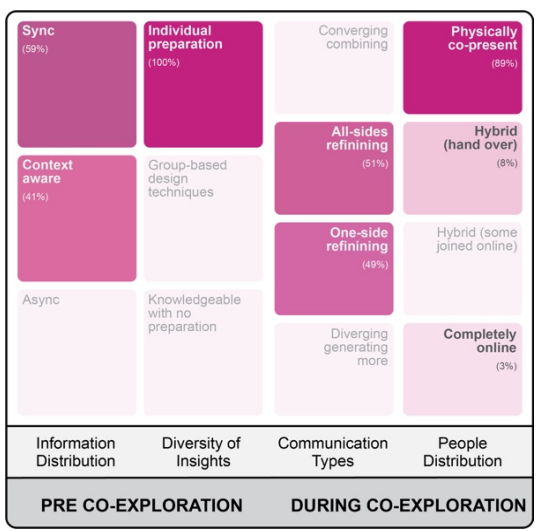

Figure 10. Co-exploration when refining initial concepts
.

Episode 3: *"Prototyping five different concepts"*

In week two, Team 5 turned sketches into lo-fi prototypes, spending a day together at one table. T5P2 explained, *"We sketched our ideas, mostly just for us to communicate different kinds of thoughts. But while we were making, we used rapid prototyping as a way to generate ideas and communicate with each other visually."* T5P3 added, *"Before this, we were only sketching and only have very vague ideas."* T5P1 further emphasized the need to *"show how it will work."*

This pattern appeared 37 times and revolved around refining ideas, either all-sides refining (19 instances) and one-side refining (18 instances). In all cases, participants came prepared, with information either pre-shared or contextually understood by the group. All sides refining involved the entire team collaboratively developing initial concepts, as seen in Episode 3. One-side refining typically was driven by external input, individuals such as studio coaches, user testers, or interviewees, provided powerful insights for overcoming complex design challenges. As observed in eight cases across six groups, this injection of an outside perspective often acted as a critical catalyst, enriching the design process by unlocking new possibilities that the team alone had not considered.

### Co-exploration when generating more ideas

Episode 4: *"First concepting meeting: 6-3-5 technique"*

In week three, Team 15 conducted their first brainstorming session. T15P3 described, *"We all had a sheet of paper and wrote 5 ideas each, and maybe sketches. We then passed the paper on and had the others build on those ideas. We did this until all of us wrote in that piece of paper."* T15P1 expressed a strong sense of co-exploration, saying, *"When we exchanged the papers, I really felt like we had co-exploration. Because we were building on each other's ideas and really getting that collective intelligence that previously wasn't there."*

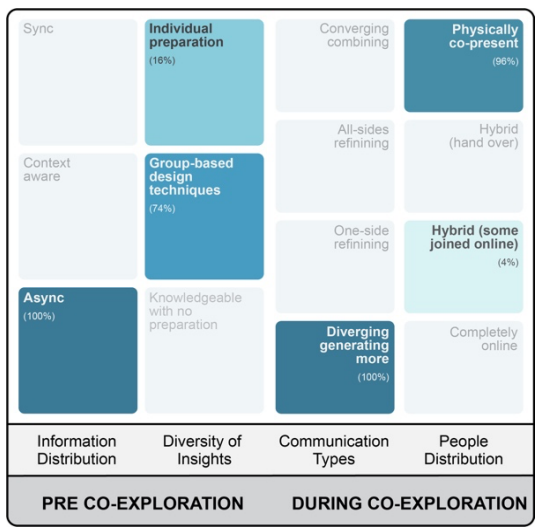

Figure 11. Co-exploration when generating more ideas

This pattern, seen in 23 activities, typically occurred during early project stages. One characteristic is that no information was shared in advance. During co-exploration, the process either began with an individual's idea serving as a trigger or entirely from scratch. Teams relied on creative techniques such as 6-3-5, crazy 8, and role-playing to generate a wide range of ideas. Although participants did not emphasize this pattern as highly impactful during final reflections, it played a critical role in fostering creativity and embracing a diverging phase, where every idea contributed to exploring possibilities.

### Co-exploration when seeking cooperation

Episode 5: *"chance meeting"*

Team 7 shared the topic with another two undergraduate students in the design studio. An impromptu meeting led to co-exploration. T7P3 described how spontaneous encounter sparked a co-exploratory process where they altered concepts based on each other's ideas: *"It was really co-exploration [...] We had to alter our concepts, and they had to alter theirs as well."* T7P2 also viewed this encounter as a co-exploration, explaining, *"We essentially built up the borders of the projects [...] So we explored together, and that exploration was based on our concepts. We also gave each other tips on how we can do it."*

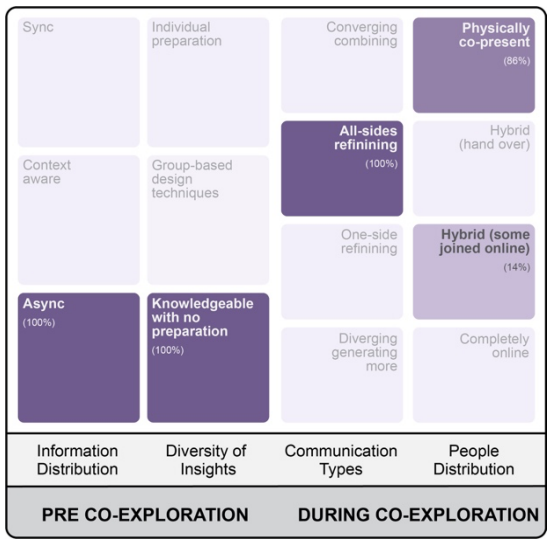

This pattern highlights the unique value of spontaneous encounters, particularly when teams work on interconnected challenges. Observed exclusively in the Design Studio 2 context, where a distributed design challenge required teams to ensure their work could integrate, this form of co-exploration was almost always unplanned. As seen in Episode 5, teams often entered these interactions without a clear agenda, only to discover unexpected cooperation opportunities. This type of co-exploration required no specific preparation and allowed projects to enrich one another while remaining distinct, hence, the communication type is converging: all refining.

Figure 12. Co-exploration when seeking cooperation

### *Different circumstances of co-exploration*

Our findings revealed the varied circumstances under which co-exploration occurs. Out of 161 recorded co-exploration activities, the vast majority (147) happened with all team members physically co-present, indicating the critical role of co-presence in the collaborative design process.

Nine co-explorations occurred in hybrid settings, where some team members joined online. For example, Team 11 held three hybrid meetings, starting with the presentations of individual assignments before engaging in co-exploration. However, members who joined remotely often felt less involved. T11P1 expressed, *"I was online, and I couldn't really always see what they were doing,"* while T5P1 echoed, *"I was writing everything down, I could hear everything, but I was not that much involved in the whole discussion."*

Two co-explorations were conducted entirely online. In one case, Team 7 divided their design topic into sub-topics and prepared individually before an online meeting using Miro and Teams. T7P2 remarked, *"I think we're probably the weirdest group when it [co-exploration] comes to the Miro discussion [...] Actually, this online meeting worked better than on-site."* When asked about the reason, T7P3 explained, *"I think we still have a little trouble getting the communication going when we are on-site,"* with T7P1 adding, *"I think we are better at writing and drawing than speaking."*

Additionally, three unique "hand-over" co-explorations were observed, where one team member acted as an information carrier to synchronize knowledge across stages. For example, in Team 16's 3D printing task, P3 shared details of previous failed attempts, enabling P2 and P4 to adjust settings and avoid the same errors. Similar hand-over co-exploration activities were observed in Teams 3 and Team 9, where members contributed insights at different stages, considering the entire process as a continuous co-exploration to achieve a shared goal.

***Active co-exploration suggests thriving design teams***

We first examined whether different levels of experience, represented in our study by the academic level that the students were enrolled in, affected engagement in co-exploration activities. A one-way ANOVA revealed no significant difference between the two groups in the number of activities (F=0.009, p=0.926) or the number of distinct patterns covered (F=0.326, p=0.577). The results indicate that regardless of academic level, both Bachelor's and Master's teams engaged in a similar number of activities and a comparable range of patterns.

We collected information on participants' final assessments: 13 teams passed the course with one of them receiving an excellence remark, and three teams did not pass. We examined the relationship between the descriptive data of co-exploration and these final assessments. Based on both their final assessments and our observations, we categorized the teams into two groups: thriving teams and struggling teams. The selection criteria for struggling teams were as follows:

1) The team experienced persistent group conflicts that hindered future collaboration (T12).
2) Teams displayed a consistent lack of interest in the design topic, which remained unchanged throughout the project (T1, T16)
3) Teams encountered prolonged unsolved problems in the design process, causing delays in subsequent design stages (T3, T4, T7, T9, T10).

Using this categorization, Figure 13 was created to depict the frequency and diversity of co-exploration activities in both thriving and struggling teams. The grey bar represents the overall frequency, while the colored segments indicate different co-exploration patterns, aligning with our previously identified five patterns. As illustrated in Figure 13, thriving teams demonstrated a higher frequency and greater diversity of co-exploration activities compared to struggling teams. Based on these observations, we proceeded with a quantitative analysis using SPSS to validate our findings. Our analysis showed that thriving teams (N=8) had a significantly higher mean frequency of co-exploration activities (M=12.75, SD=3.012) than struggling teams (N=8, M=7.38, SD=2.200). An independent samples t-test confirmed this difference. The distributions of both groups were sufficiently normal for the t-test, and Levene's test confirmed the homogeneity of variances, F=0.79, p=0.389. The t-test revealed a statistically significant difference, t=4.08, p=0.001, indicating that thriving teams engaged in co-exploration activities significantly more frequently than struggling teams. Additionally, we found that thriving teams (N=8) exhibited a significantly greater diversity of co-exploration patterns (M=4.38, SD=0.518) compared to struggling teams (N=8, M=3.25, SD=1.053). Another independent sample t-test showed a statistically significant effect, t=2.75, p=0.01, confirming that thriving teams also experienced a higher diversity of co-exploration patterns.

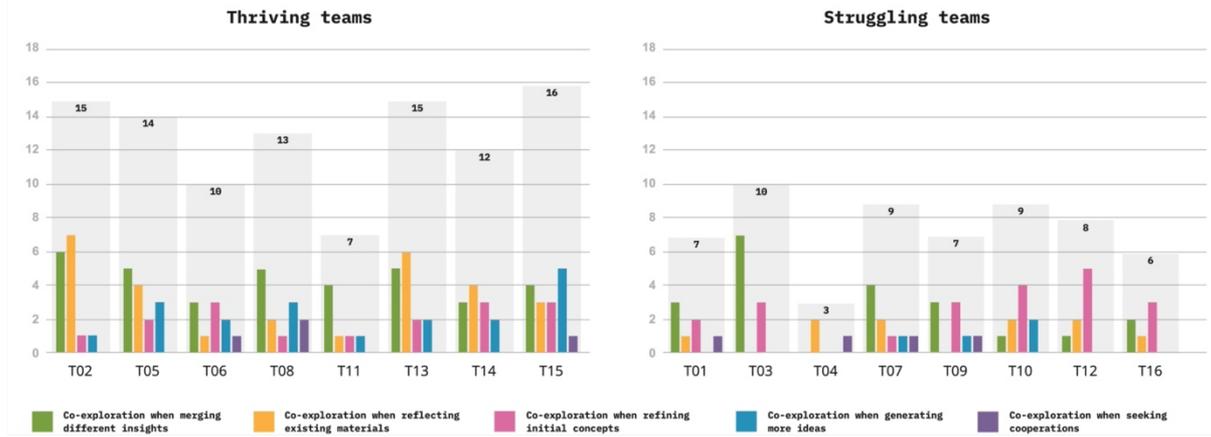

Figure 13. Pattern distribution in thriving and struggling teams. The grey bar in the figure represents the frequency, with different colors of small bars indicating various patterns of co-exploration, consistent with our previous results on the five patterns.

## Discussion

Throughout our research, the progression from identifying broad features to developing an analytical framework and ultimately recognizing distinct patterns highlighted the complexity of co-exploration. Each stage of analysis contributed to a deeper and more layered understanding. Here, we discuss the findings and implications related to what, where, and why of co-exploration within collaborative design processes.

### *Co-exploration as an emergent experience*

Among 1,562 recorded activities, only 220 were identified as co-exploration, suggesting that structural similarities alone do not guarantee its emergence. For example, making a mind map is an activity, but the act of making a mind map does not necessarily lead to co-exploration. The same activity may foster co-exploration in one context but merely decision-making in another, as the experience depends on how participants engage with one another. This suggests that co-exploration is defined not by the activity , but by the quality of interaction that unfolds within it. From this perspective, we distinguish between collaborative activities and the shared experience of co-exploration: while co-exploration arises within activities, it is, at its core, a shared experience among the individuals who take part.

Our analysis of the five identified patterns, supported by detailed episodes, further reinforces that co-exploration is emergent: it does not automatically arise in collaborative design but depends on specific conditions. These conditions correspond to the four dimensions of our analytical framework: (1) how information is shared in advance, (2) how diverse insights are introduced, (3) the type of communication that occurs, and (4) how participants are physically or virtually distributed.

Recognizing co-exploration as an emergent experience shifts the focus from simply selecting the "right" activity to shaping the conditions under which it is more likely to occur. This raises an important question for future research and practice: how might we intentionally design for these conditions to make certain forms of co-exploration more likely to emerge?

***Co-exploration occurs naturally***

Co-exploration was observed across design teams of different academic levels and in both thriving and struggling teams. While struggling teams engaged in fewer instances, they still experienced moments of exchanging knowledge, building on each other's ideas, and jointly exploring problem and solution spaces. This suggests that co-exploration is not dependent on expertise or overall team success but remains a fundamental aspect of collaborative design. Notably, we found no significant difference between master's and bachelor's teams in the number or diversity of co-exploration activities.

Two factors may explain this. First, co-exploration can occur across a wide range of activities and with varying intensity. Our data allowed us to identify and categorize co-exploration but did not capture differences in the depth or quality of the experience between groups. Second, because co-exploration is not formally introduced or taught as a concept, instructors do not explicitly guide or influence its occurrence. Instead, it appears to emerge organically from the design process and the instincts of the designers themselves.

***Co-exploration is more than diverging***

As introduced earlier in this paper, the design process is iterative and exploratory, evolving through dynamic interactions between problem and solution spaces, an approach often characterized as "design as exploration[45]". In the design field, exploration is frequently associated with divergence, meaning an intentional expansion of perspectives to generate more alternatives[46]. When this process occurs in design collaboration, individual exploration extends into co-exploration, often leading to the assumption that co-exploration simply involves more people contributing to an expanded pool of possibilities. Brainstorming, for example, is a well-known designerly activity that embodies this assumption. It brings people together, fostering creativity to open up possibilities, uncover hidden issues, and reframe problems, all of which inherently require divergence. However, our findings reveal that while co-exploration includes divergence, it is not solely a divergent process. It also involves refining ideas, reflecting on insights, converging toward solutions, and coordinating efforts. These different patterns of co-exploration unfold throughout the entire design process, serving purposes more than merely generating alternatives. This challenges the traditional view of co-exploration as primarily divergent, revealing it instead as a multifaceted and dynamic experience that unfolds through a variety of collaborative activities.

***Co-exploration is more than the sum of individual exploration***

As one participant noted, *"the end result is more than the sum of individual capabilities"* (T8P4). This sentiment, echoed by others who described a *"collective intelligence"* (T15P1), points directly to the emergence of collective creativity. This is not merely an additive process of combining individual skills; it is a generative one where co-exploration fosters more ideas to emerge. Our findings suggest that co-exploration acts as an engine that drives this process at a group level: during these activities, teams exchange perspectives, refine ideas, and challenge assumptions, producing insights that no single member could have generated alone.

This aligns with research on collective intelligence, which finds that a team's success depends not just on individual intelligence but on how well members listen, communicate, and build on each other's ideas[47].

The true value of the "co-" in co-exploration, therefore, is its power to transform individual knowledge into collective creativity through shared, interactive experience. The reported failure of remote collaboration provides compelling evidence for this argument[48]: teams found that simply increasing individual exploration and adding group discussions did not compensate for the loss of co-exploration in remote collaboration. This presents a critical challenge and opportunity for future research. If co-exploration plays role of unlocking a team's creative potential, how can we design supporting tools that move beyond simple information exchange in remote design collaboration? How can technology support the lived, shared experience of co-exploration, fostering the togetherness and interactive synergy required to the collaborative design processes?

### The sense of togetherness

From the group reflection sessions, one of the five key features of co-exploration was the importance of being "together," both in physical co-presence and in a shared sense of connection when apart. Examining all 161 recorded co-exploration activities, we found that the vast majority (147) occurred in physically co-present settings, suggesting a strong preference for face-to-face collaboration. The remainder took place in fully online, hybrid, or, in three cases, handover.

Interestingly, none of the non-co-present co-exploration activities occurred at the start of the collaborative process. A likely reason is that creativity is fostered when there is no fear of judgment or external evaluation; people need to feel safe to share half-formed, unconventional, or "silly" ideas, which are often the seeds of breakthrough concepts [49]. Early co-present interactions may therefore build *we-experiences* fostered a sense of *habitual togetherness*[50] that allow co-exploration to continue remotely later in the process.

However, participants also reported challenges in remote co-exploration, such as difficulty viewing teammates' work (T11P1) or feeling reduced to a passive role, like notetaking (T5P1). These barriers may help explain the relatively low occurrence of non-co-present co-exploration. At the same time, they highlight opportunities for future research: to design tools and practices that better support visibility, active participation, and shared presence in remote settings, thereby enhancing the conditions for co-exploration beyond physical co-location.

### *Synchronizing or contextualizing*

In collaborative design, gathering and sharing information from diverse sources are crucial[51]. Particularly in remote collaboration, the availability and accessibility of data are vital for confidence in the collaborative process[52]. Our study, however, reveals a more nuanced picture of information sharing prior to co-exploration activities. We found that only 48 co-exploration activities involved fully synchronized pre-sharing of information, while 89 relied on asynchronous sharing, and 24 were based solely on contextual awareness without comprehensive pre-sharing. This suggests that complete information sharing is not always essential for effective co-exploration.

Two reasons may explain the lower rate of pre-sharing observed in our study. First, the educational context might have encouraged participants to trust in their ability to align and share information during scheduled meetings. Second, the loosely coupled nature of tasks allowed independent preparation, with convergence occurring later in discussions. These insights align with Kvan's assertion that synchronicity is less critical for cooperative tasks, which require less coordination compared to collaboration which demand multi-level communication and structured planning[53]. Our findings suggest that designers might engage in more cooperative work than anticipated, where the immediacy of synchronous information sharing is not always a necessity.

### *The impact of co-exploration*

Our quantitative analysis (section 6.7) revealed a contrast in the frequency and diversity of co-exploration activities between thriving and struggling teams. Thriving teams engaged more frequently in co-exploration activities and exhibited a broader range of patterns while struggling teams participated less and demonstrated fewer types of patterns. The success of design outcomes is complex and influenced by a multitude of factors, such as the nature of design challenge[54], group culture[55], communication dynamics[56], team relationships and leadership styles[57]. While it is too early to establish a direct causal relationship between co-exploration and successful design outcomes, the observed differences in co-exploration activity suggest its potential importance. Thus, we argue that co-exploration should be considered a contributing element that influences the trajectory of design success, warranting further investigation by design researchers.

*Limitations and future work*

We acknowledge several limitations in our study that should be addressed in future research. First, there is a possibility that the first author's involvement might have influenced participants' behaviors. Knowing that their design processes were under investigation and that the researcher aimed to uncover certain insights, participants might have modified their actions or presented their work in a more favorable light to align with perceived research expectations[58]. To mitigate the potential bias, we conducted a long-term study and explicitly informed participants that the first author had no role in their final assessment, assuring them that no details would be shared with their examiners.

Second, the scope and diversity of our sample may limit generalizing our results to all collaborative design processes. Our participants were design students, whose skills may differ from those of professional designers. Moreover, the final assessments considered their learning progress, which might not fully reflect the quality of the design outcomes. To address this limitation, future work will include consultations with design experts to evaluate the applicability of our findings beyond the educational context.

Additionally, we observed that participants sometimes used our weekly interviews as opportunities for their group reflection sessions. This had a dual impact: On the positive side, it demonstrated that participants valued these weekly reflections, possibly enhancing their motivation to engage with our study. However, it also raises the possibility that our research may have influenced their processes, potentially altering the frequency or nature of their group reflection practices. While we recognize this limitation, we believe it did not fundamentally change the essence of the teams' co-exploration activities as the diary-interviews did not offer new co-exploration techniques, nor did they offer insights into co-exploration mechanisms.

## Conclusion

This paper presented a five-month longitudinal observational study of co-exploration within collaborative design processes, involving 61 students across 16 design teams. Through the analysis of group reflection sessions and iterative coding, we identified five features of co-exploration that shaped our four-dimensional analytical framework: information distribution, diversity of insights, communication types, and people distribution. This framework allowed us to categorize co-exploration experiences and recognize distinct patterns, offering a deeper understanding of its role in collaborative design.

Our findings show that co-exploration is not confined to the early, divergent phases of design. It also encompasses reflection, refinement, convergence, and coordination throughout the process. Crucially, the "co-" in co-exploration lies in its ability to transform individual knowledge into collective creativity through a shared, interactive experience. Togetherness, both as a practical condition and as a social value, plays an important role in enabling this transformation. As a result, teams achieve outcomes exceeding what individuals could achieve alone. Notably, thriving teams engaged in co-exploration more frequently and demonstrated a broader range of patterns than struggling teams. While our statistical analysis revealed significant differences between these groups, we do not assert that increased co-exploration directly leads to better design outcomes. Instead, we suggest that the presence and diversity of co-exploration activities can serve as indicators of team dynamics, creative engagement, and the health of the collaborative process.

Based on these insights, we propose the following recommendations for designers and facilitators of collaborative projects:

(1) **Foster shared understanding early:** In early design stages, co-exploration should prioritize individual preparation and open discussion to create shared understanding, even if tangible results aren't achieved. Designers should avoid rushing to merging ideas and instead focus on collective ideation and establishing a common ground.

(2) **Address ambiguity:** When feeling lost or dealing with ambiguous issues, co-exploration often occurs when *reflecting on existing materials*, prompting teams to revisit their journey and previous collaborations to regain clarity. Using techniques such as role-playing, rapid prototyping, or mind mapping, combined with open discussions based on existing materials and knowledge, can help address challenges and clarify objectives.

(3) **Incorporate external input:** Engaging external stakeholders, such as coaches, peers, or industry experts, can bring fresh perspectives and enrich the co-exploration process, especially when teams face complex design challenges (as seen in *Co-exploration when refining initial concepts*).

(4) **Adapt Dynamically as a Team:** Recognizing and adapting to each team's unique working style is essential. The responsibility for adapting to a team's unique working style should be shared by all members. This collective awareness involves tailoring collaboration settings to the team's needs, but also consciously shifting between different patterns of co-exploration.

(5) **Value spontaneous encounters:** In co-exploration when seeking cooperation, unplanned and impromptu discussions often arise when design teams work on interconnected challenges, sparking collaboration and leading to unexpected opportunities that benefit all involved. Fostering an environment where such organic interactions are encouraged can enhance the collaborative design process.

In sum, we have shown that co-exploration is a multifaceted and emergent experience that moves beyond divergence and the aggregation of individual ideas, turning collaboration into a process that builds collective intelligence, deepens togetherness, and strengthens the conditions for a creative, healthy, and thriving design practice.

# Appendix

Table 1. Overview of collected data

| Team | Times of weekly interview | Total duration of weekly interview recordings | Numbers of design activities | Numbers of collected images | Numbers of co-exploration | Numbers of selected cases of co-exploration |
|---|---|---|---|---|---|---|
| 1 | 9 | 3h 28m 24s | 99 | 47 | 11 | 7 |
| 2 | 11 | 4h 46m 30s | 82 | 60 | 21 | 15 |
| 3 | 10 | 3h 54m 17s | 107 | 97 | 15 | 10 |
| 4 | 10 | 4h 4m 24s | 50 | 8 | 3 | 3 |
| 5 | 10 | 6h 6s | 191 | 136 | 18 | 14 |
| 6 | 10 | 6h 24m 35s | 112 | 101 | 12 | 10 |
| 7 | 11 | 3h 34m 22s | 70 | 51 | 11 | 9 |
| 8 | 10 | 5h 46m 43s | 80 | 122 | 18 | 13 |
| 9 | 10 | 4h 41m 28s | 96 | 53 | 10 | 7 |
| 10 | 10 | 3h 41m 29s | 69 | 32 | 12 | 9 |
| 11 | 10 | 5h 39m 43s | 109 | 85 | 8 | 7 |
| 12 | 11 | 4h 59m 18s | 103 | 55 | 12 | 8 |
| 13 | 10 | 4h 49m 39s | 116 | 47 | 19 | 15 |
| 14 | 10 | 4h 29m 15s | 66 | 56 | 13 | 12 |
| 15 | 10 | 9h 3m 5s | 124 | 145 | 28 | 16 |
| 16 | 9 | 3h 26m 44s | 88 | 68 | 9 | 6 |
| total | 161 | 78h 50m 2s | 1562 | 1163 | 220 | 161 |